\documentclass[onecolumn,amsmath,amssymb]{qm2008_abs}
\usepackage{graphicx}% Include figure files
\usepackage{dcolumn}% Align table columns on decimal point
\usepackage{bm}% bold math
\usepackage{color}

\newcommand{\dd}{ {\textrm d}}
\begin{document}

\title{{\Large Cold Nuclear Modifications at  RHIC and LHC }}% Force line breaks with \\
%%%%%%%%%%%%%%%%%%%%%%%%%%%%%%%%%%%%%%%%%%%%%%%%%%%%%%%%%%%%%%%%%%%%%%%

\bigskip
\bigskip
\author{\large G.G. Barnaf\"oldi\textsuperscript{1,2}}
\email{bgergely@rmki.kfki.hu}

\author{\large G. Fai\textsuperscript{1}}

\author{\large P. L\'evai\textsuperscript{2}}

\author{\large B. A. Cole\textsuperscript{3}}

\author{\large G. Papp\textsuperscript{4}}
\bigskip
\bigskip

\affiliation{\textsuperscript{1}Center for Nuclear Research, Kent State 
University, Kent, OH 44242, USA}

\affiliation{\textsuperscript{2}MTA KFKI RMKI, Research Institute for 
Particle and Nuclear Physics, P.O. Box 49, Budapest 1595, Hungary}

\affiliation{\textsuperscript{3}Nevis Laboratory, Columbia University, 
New York, NY, USA}

\affiliation{\textsuperscript{4}Dept. for Theoretical Physics, 
E\"otv\"os University, P\'azm\'any P. s\'et\'any 1/A, Budapest 1117, Hungary}
\bigskip
\bigskip

%%%%%%%%%%%%%%%%%%%%%%%%%%%%%%%%%%%%%%%%%%%%%%%%%%%%%%%%%%%%%%%%%%%%%%%
\begin{abstract}
\leftskip1.0cm
\rightskip1.0cm
We use recent nuclear parton distributions, among them the
Hirai\,--\,Kumano\,--\,Nagai (HKN) and Eskola\,--\,Paukkunen\,--\,Salgado 
(EPS08) parameterizations, in our pQCD-improved parton model to calculate
the nuclear modification factor, $R_{AA'}(p_T)$, at RHIC and at the LHC. 
At RHIC, the deuteron-gold nuclear modification factor for pions, measured at 
$p_T \geq 10$~GeV/c in central collisions, appears to deviate more from unity 
than the model results. The slopes of the calculated 
$R_{dAu}(p_T)$ are similar to the slopes of the PHENIX pion and photon data. 
At LHC, without final-state effects we see a small enhancement of 
$R_{dPb}(p_T)$ in the transverse momentum range $10$ GeV/c 
$ \geq p_T \geq 100$~GeV/c for most parameterizations. The inclusion of 
final-state energy loss will reduce the $R_{dPb}(p_T)$ values.
 
\end{abstract}
%%%%%%%%%%%%%%%%%%%%%%%%%%%%%%%%%%%%%%%%%%%%%%%%%%%%%%%%%%%%%%%%%%%%%%%
\maketitle
%%%%%%%%%%%%%%%%%%%%%%%%%%%%%%%%%%%%%%%%%%%%%%%%%%%%%%%%%%%%%%%%%%%%%%%
\section{Introduction}

Nuclear modifications in the physics of partonic degrees of freedom
arise either due to the environment in the initial nuclei around the 
incoming parton or from the interaction with the partonic matter created 
in the final state of a nuclear collision. Coherent quantum 
effects may connect and mix initial and final state phenomena. Here we 
investigate '{\it cold nuclear effects}' defined as effects that do 
not originate in connection with the thermalized partonic 
matter in collisions between two heavy nuclei~\cite{Cole:2007,GGB:QM}.        

Cold nuclear effects are usually described as modifications of the 
parton distribution functions (PDF) into nuclear parton distributions 
(nPDF). At sufficiently small momentum fraction, this has been referred 
to as nuclear shadowing, but further effects, like nuclear multiple 
scattering, EMC, higher-twist effects, etc. need to 
be included in a more complete description. In this short review 
we present the high-$p_T$ and high-$x_T$ behavior of some of these 
modifications, using shadowing/nuclear PDF parameterizations by
HIJING~\cite{HIJING}, EKS~\cite{EKS}, EPS08~\cite{EPS08}, and 
HKN~\cite{HKN}. 

At present, cold nuclear modifications can be tested by experimental data 
at RHIC energies on hadron spectra from $dAu$ collisions, or by direct 
photon spectra from $dAu$ or $AuAu$ collisions, which are essentially not 
influenced by non-Abelian jet energy loss. Note that recent theoretical 
studies point out the possibility of a small energy loss in $pA$ or 
$dA$ type collisions~\cite{Cole:2007,VitevCold}. 

%%%%%%%%%%%%%%%%%%%%%%%%%%%%%%%%%%%%%%%%%%%%%%%%%%%%%%%%%%%%%%%%%%%%%%%
\section{Cold Nuclear Effects at RHIC}

\begin{figure}[ht]
\begin{center}
\resizebox{70mm}{50mm}{\includegraphics{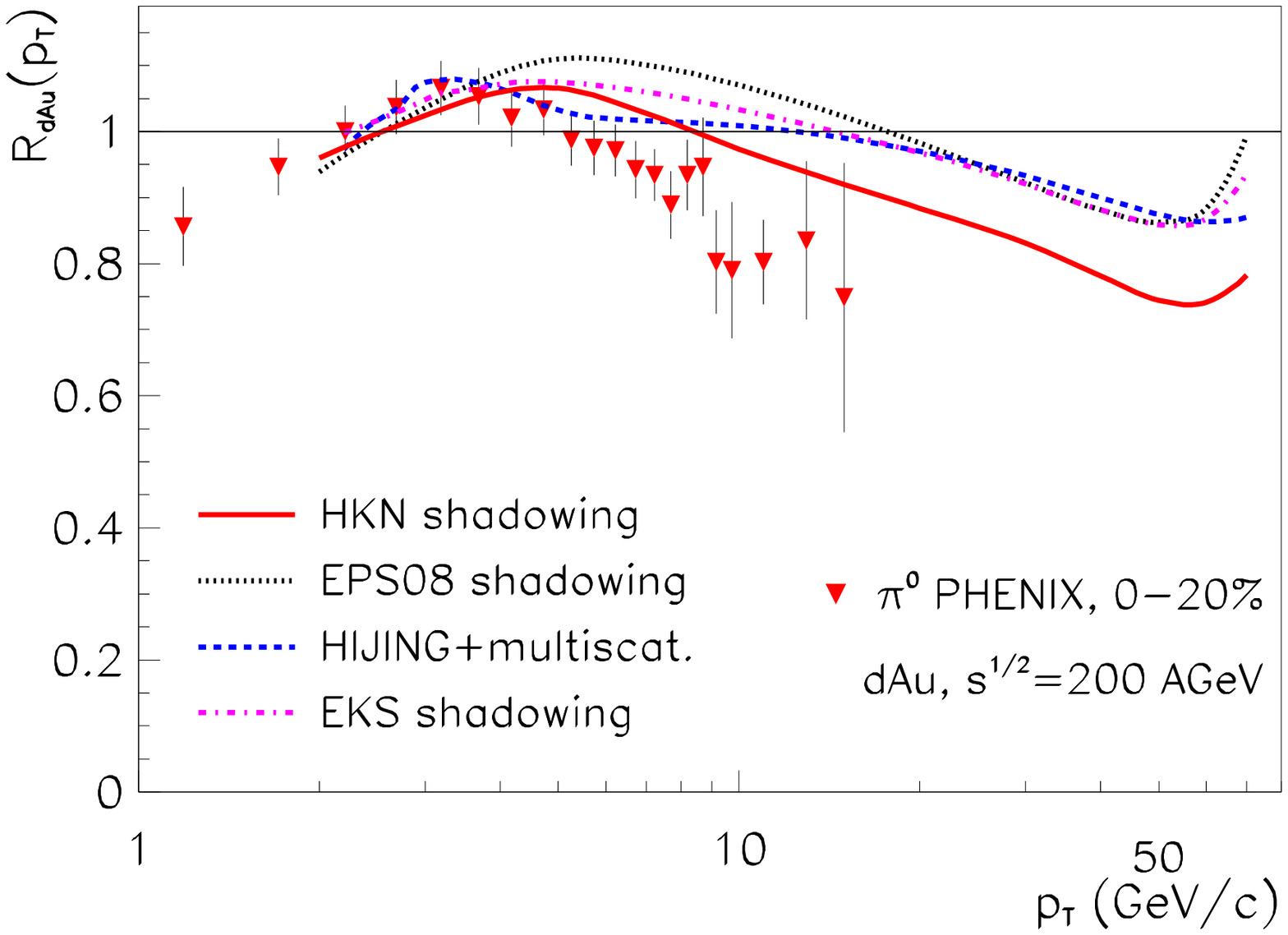}}
%\resizebox{70mm}{50mm}{\includegraphics{qmslope.eps}}
\resizebox{70mm}{50mm}{\includegraphics{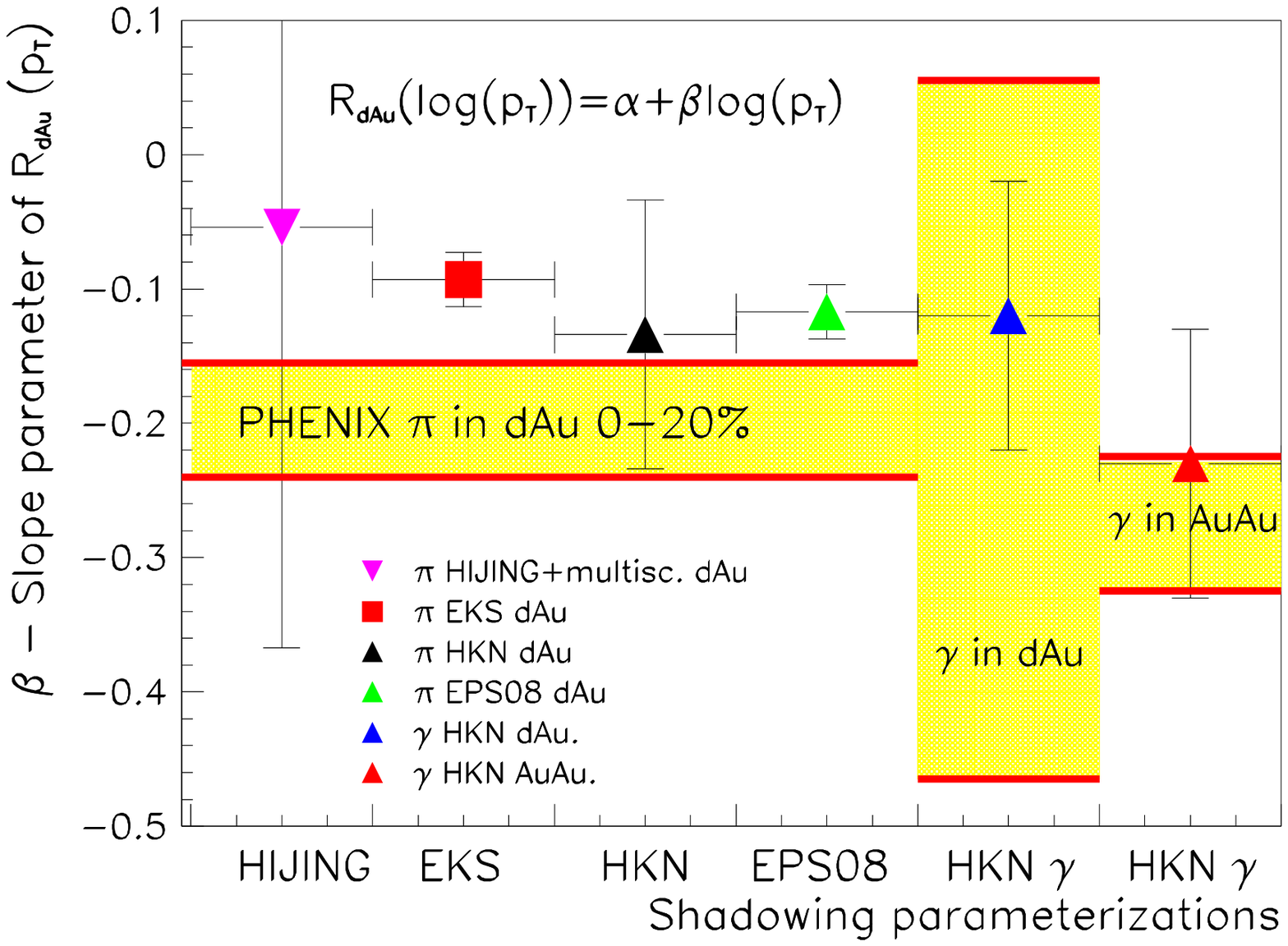}}
\caption{Calculated $R^{\pi}_{dAu}(p_T)$ compared to most central 
PHENIX data~\cite{PHENIX} ({\sl left panel}), and the extracted slope  
parameter, $\beta$ of Eq.(\ref{beta}) in different models 
({\sl right panel}) and Ref.s~\cite{PHENIX,3,4}.}
\label{fig:1} 
\end{center}
\end{figure}

We calculated the nuclear modifications at RHIC energies up to 
$p_T \lesssim 70$ GeV/c using various shadowing parameterizations in 
our pQCD improved parton model~\cite{GGB:HQ}. Results are compared to 
the preliminary PHENIX data up to the highest measured momenta 
($p_T \lesssim 18$ GeV/c)~\cite{PHENIX,3,4}. On the left side of 
Fig.~\ref{fig:1}, 
$R^{\pi}_{dAu}(p_T)$ is plotted for neutral pions at $\sqrt{s}=200$ $A$GeV 
against a logarithmic $p_T$ scale, using the HIJING, EKS, EPS08, and HKN 
parameterizations. The HIJING shadowing is understood with its 
accompanying multiple scattering~\cite{GGB:HQ}. 

\vspace*{-0.3truecm}

All calculated $R^{\pi}_{dAu}(p_T)$ curves have an approximately linear 
dependence on $\log(p_T)$ in most of the EMC region ($15$ GeV/c 
$\lesssim p_T \lesssim 50$ GeV/c). Thus, we used a simple linear 
function of $\log(p_T)$ to get the slope parameter $\beta$:
\begin{equation}
R_{dAu}(\log(p_T))=\alpha + \beta\cdot\log(p_T)  \,\,\,  .
\label{beta}
\end{equation}
The slope $\beta$ is plotted in the right panel of Fig.~\ref{fig:1}.
We use the best-fitted linear curve in the above $p_T$ interval to compare 
the calculated slopes to the ones extracted from the data. The slopes 
of the model curves are quite similar, although they 
are smaller (in absolute value) than the measured quantities.
This may be due to final state effects~\cite{Cole:2007} 
or isospin effects~\cite{Isospin}. In addition to pion production slopes
in $dAu$ collisions, direct photon production in both $dAu$ and $AuAu$ 
collisions is plotted. Our results are compared to the preliminary 
PHENIX data~\cite{PHENIX,3,4}. These data have decreasing tendencies with 
$p_T$, leading to negative $\beta$ values similar to those 
in $\pi^0$ production. In the case of the $dAu$ photon data, the 
large error bars manifest themselves in a large uncertainty
of the extracted slope. Direct $\gamma$ production in 
$AuAu$ shows a stronger decrease, since the initial state 
nuclear effects enter the convolution integral twice in this case. 

We analyzed the theoretical uncertainties in our 
model~\cite{Cole:2007,GGB:QM}. We found that at  $p_T > 3$ GeV/c our 
model is not sensitive to the scale choice. Nuclear PDF 
parameterizations vary both the position of the Cronin peak 
and the high-$p_T$ slope of $R_{dAu}(p_T)$. The HKN parameterization 
allowed us to analyze the errors via the Hessian method~\cite{HKN}. An 
almost constant $\pm 10\%$ uncertainty was found in the whole 
transverse momentum range.  

It can be seen from Fig.~\ref{fig:1} that the EPS08 and HKN nPDF 
parameterizations give slopes close to the experimental ones. However, 
HKN comes closer to the measured points at high $p_T$ (see the left 
panel). Note that the agreement with the data can be improved by the 
introduction of a small opacity, as we have done in Ref.~\cite{Cole:2007}. 
       
%%%%%%%%%%%%%%%%%%%%%%%%%%%%%%%%%%%%%%%%%%%%%%%%%%%%%%%%%%%%%%%%%%%%%%%
\section{Predictions for the LHC}

\begin{figure}[ht]
\begin{center}
\resizebox{\textwidth}{80mm}{\includegraphics{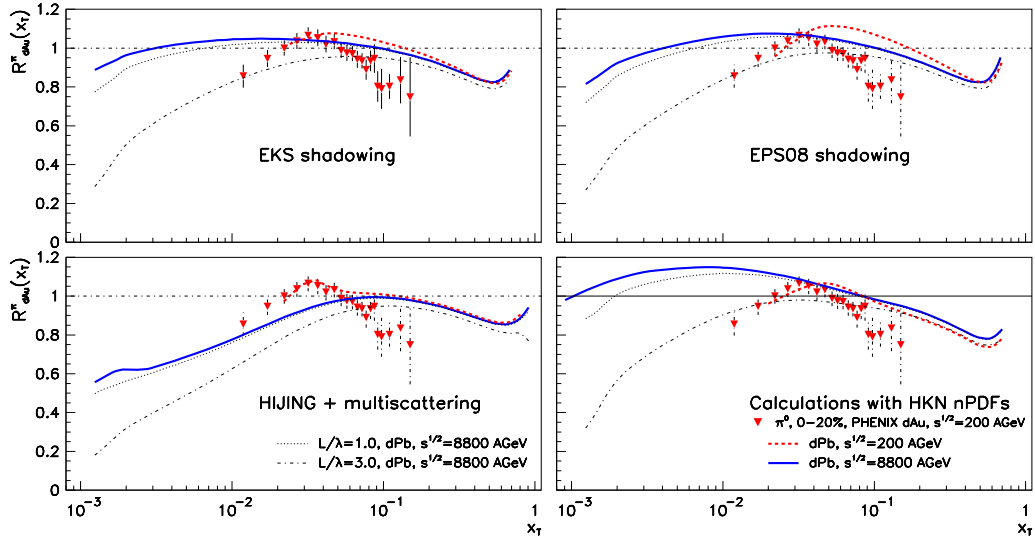}}
\caption{Nuclear modifications $R^{\pi}_{dA}(x_T)$ at RHIC 
({\sl dashed}) and LHC ({\sl solid}) energies with different shadowing 
parameterizations. Preliminary PHENIX data are plotted as 
functions of $x_T$. The effect of cold energy loss is indicated by
{\sl thin dotted} and {\sl dot-dashed lines} for $L/\lambda_g=1$ and 
$3$, respectively.}
\label{fig:2} 
\end{center}
\end{figure}

Based on Ref.s~\cite{Cole:2007,GGB:QM,E706} we estimated the cold 
nuclear modifications at LHC energies with larger intrinsic transverse 
momenta. We apply the same parameterizations used above. 
In Fig.~\ref{fig:2} we plot $R^{\pi}_{dA}(x_T)$ in $dPb$ collisions
at midrapidity at RHIC and LHC energies as functions of transverse
momentum fraction, $x_T = 2 p_T / \sqrt{s}$. The preliminary PHENIX 
$dAu$ data on neutral pion production are also shown. The different 
panels display different shadowing parameterizations as noted on  
Fig.~\ref{fig:2}.  

Though at RHIC energies all parameterization give similar results, at 
LHC we see that at $x_T \approx 3 \cdot 10^{-3}$ HIJING yields a strong 
($\approx  30\%$) suppression, while HKN gives a $10\%$ enhancement at 
the same $x_T$ values. The EKS and EPS08 results predict a $5-10\%$ 
enhancement (suppression) around $x_T \approx 10^{-2}$ ($10^{-3}$). The 
EPS08 results show somewhat stronger suppression than EKS at low $x_T$,
due to the inclusion of high-rapidity RHIC data in the fitting procedure 
of this recent parameterization. For the high momentum-fraction region 
both of these models show a similar behavior due to the EMC effect. 
HIJING and HKN have a reasonable agreement between RHIC and LHC energies
(they scale with $x_T$), while EKS and especially EPS08 display
less precise scaling at around the anti-shadowing maximum.

\vspace*{-0.3truecm}

Finally we turn to possible cold energy loss at the LHC. The standard
estimate is $\dd N/ \dd y\approx 1500-4000$ in central 
$PbPb$ collisions at LHC energies~\cite{LastCall}. While in 
central $dPb$ collisions the geometrical cross section is small, 
using $L/\lambda \sim \langle N_{part}\rangle ^{1/3}$ from~\cite{GGB:HQ}  
one obtains $\dd N/ \dd y\approx 500-2000$, still large, in central $dPb$
collisions. This motivates us to examine the effect of a small 
energy loss in the final state. We apply opacity values 
$L/\lambda_g \lesssim 3$ for this illustration. The results are indicated 
as thin dotted ($L/\lambda_g=1$) and dot-dashed ($L/\lambda_g=3$) lines in 
Fig.~\ref{fig:2}. 

The energy loss is stronger at the lower $x_T$ values. Here 
the relative particle yield is suppressed, and $R_{dA}(x_T)$ is 
steeper. In the high $p_T$ region the suppression loses strength as
it increases with $\sim \log(E)$. We tested this effect at RHIC and 
LHC energies in $AuAu$ ($PbPb$) collisions~\cite{GGB:QM08}.

%%%%%%%%%%%%%%%%%%%%%%%%%%%%%%%%%%%%%%%%%%%%%%%%%%%%%%%%%%%%%%%%%%%%%%%
\section{Summary}

We analyzed the high transverse-momentum behavior of the nuclear 
modifications in $dA$ collisions. Using preliminary PHENIX data we checked
several common nuclear shadowing parameterizations.
We compared the logarithmic 
slopes of the nuclear modification factor in the $15$ GeV/c 
$\lesssim p_T \lesssim 50$ GeV/c region. The shadowing parameterizations 
investigated have an almost linear behavior with a negative slope in $dAu$
collisions at RHIC, which can be attributed to the EMC effect. While we 
found all
studied shadowing models reliable at intermediate $p_T$, at high 
transverse momenta the EPS08 and the HKN nuclear PDF seem to give the best 
agreement with the PHENIX data. We tested the $x_T$ scaling of midrapidity 
$\pi^0$ production in $dA$ collisions at RHIC and LHC energies in the models.
At high $x_T$ most models have a similar behavior, but EPS08 deviates from 
precise scaling around $x_T\approx 6\cdot 10^{-2}$. At lower $x_T$, 
$R_{dA}(x_T)$ shows a $5-15\%$ enhancement in the HKN, EKS, and EPS08 
models. The
enhancement is absent in the HIJING with multiscattering model. 
We tested the effect of a small energy loss, which counteracts 
any enhancement in $dPb$ collisions and yields a  
wide-range suppression of the $R_{dA}(x_T)$. In light of these results
and open questions, we emphasize the need for
$pPb$ (or at least $dPb$) measurements at the LHC.

%%%%%%%%%%%%%%%%%%%%%%%%%%%%%%%%%%%%%%%%%%%%%%%%%%%%%%%%%%%%%%%%%%%%%%%
\section{Acknowledgments}
One of the authors (GGB) would like to thank the organizers for local
support. Our work was supported in part by Hungarian OTKA PD73596,
T047050, NK62044, and IN71374, by the U.S. Department of Energy under
grant U.S. DOE DE-FG02-86ER40251, and jointly by the U.S. and Hungary under
MTA-NSF-OTKA OISE-0435701.

%%%%%%%%%%%%%%%%%%%%%%%%%%%%%%%%%%%%%%%%%%%%%%%%%%%%%%%%%%%%%%%%%%%%%%%

%%%%%%%%%%%%%%%%%%%%%%%%%%%%%%%%%%%%%%%%%%%%%%%%%%%%%%%%%%%%%%%%%%%%%%%
\end{document}